\documentclass[twocolumn,english,aps,floatfix,showpacs,prb]{revtex4}
\usepackage[T1]{fontenc}
\usepackage[latin1]{inputenc}
\usepackage{textcomp}
\usepackage{amsmath}
\usepackage{graphicx}
\usepackage{epstopdf}
\usepackage{multirow}
\usepackage{babel}
\usepackage{babel}
\usepackage{babel}
\usepackage{babel}
\usepackage{babel}
\usepackage{babel}

\makeatletter
\@ifundefined{textcolor}{}
{
 \definecolor{BLACK}{gray}{0}
 \definecolor{WHITE}{gray}{1}
 \definecolor{RED}{rgb}{1,0,0}
 \definecolor{GREEN}{rgb}{0,1,0}
 \definecolor{BLUE}{rgb}{0,0,1}
 \definecolor{CYAN}{cmyk}{1,0,0,0}
 \definecolor{MAGENTA}{cmyk}{0,1,0,0}
 \definecolor{YELLOW}{cmyk}{0,0,1,0}
 }
\@ifundefined{textcolor}{}{
 \definecolor{BLACK}{gray}{0}
 \definecolor{WHITE}{gray}{1}
 \definecolor{RED}{rgb}{1,0,0}
 \definecolor{GREEN}{rgb}{0,1,0}
 \definecolor{BLUE}{rgb}{0,0,1}
 \definecolor{CYAN}{cmyk}{1,0,0,0}
 \definecolor{MAGENTA}{cmyk}{0,1,0,0}
 \definecolor{YELLOW}{cmyk}{0,0,1,0}
 }
\@ifundefined{textcolor}{}{
 \definecolor{BLACK}{gray}{0}
 \definecolor{WHITE}{gray}{1}
 \definecolor{RED}{rgb}{1,0,0}
 \definecolor{GREEN}{rgb}{0,1,0}
 \definecolor{BLUE}{rgb}{0,0,1}
 \definecolor{CYAN}{cmyk}{1,0,0,0}
 \definecolor{MAGENTA}{cmyk}{0,1,0,0}
 \definecolor{YELLOW}{cmyk}{0,0,1,0}
 }
\@ifundefined{textcolor}{}{
 \definecolor{BLACK}{gray}{0}
 \definecolor{WHITE}{gray}{1}
 \definecolor{RED}{rgb}{1,0,0}
 \definecolor{GREEN}{rgb}{0,1,0}
 \definecolor{BLUE}{rgb}{0,0,1}
 \definecolor{CYAN}{cmyk}{1,0,0,0}
 \definecolor{MAGENTA}{cmyk}{0,1,0,0}
 \definecolor{YELLOW}{cmyk}{0,0,1,0}
 }
\makeatother

\begin{document}

\title{Plaquette order and deconfined quantum critical point in the spin-1 bilinear-biquadratic
Heisenberg model on the honeycomb lattice}
\author{H. H. Zhao$^{1}$, Cenke Xu$^{2}$, Q. N. Chen$^{3}$, Z. C. Wei$^{1}$,
M. P. Qin$^{1}$}
\author{G. M. Zhang$^{4}$}
\email{gmzhang@tsinghua.edu.cn}
\author{T. Xiang$^{1,3}$}
\email{txiang@iphy.ac.cn}
\affiliation{$^{1}$Institute of Physics, Chinese Academy of Sciences, Beijing 100190,
China}
\affiliation{$^{2}$Department of Physics, University of California, Santa Barbara, CA
93106, USA}
\affiliation{$^{3}$Institute of Theoretical Physics, Chinese Academy of Sciences, P.O.
Box 2735, Beijing 100190, China}
\affiliation{$^{4}$State Key Laboratory of Low-Dimensional Quantum Physics and Department
of Physics, Tsinghua University, Beijing 100084, China}
\date{\today}

\begin{abstract}
We have precisely determined the ground state phase diagram of the quantum
spin-1 bilinear-biquadratic Heisenberg model on the honeycomb lattice using
the tensor renormalization group method. We find that the ferromagnetic,
ferroquadrupolar, and a large part of the antiferromagnetic phases are
stable against quantum fluctuations. However, around the phase where the
ground state is antiferro-quadrupolar ordered in the classical limit,
quantum fluctuations suppress completely all magnetic orders, leading to a
plaquette order phase which breaks the lattice symmetry but preserves the
spin SU(2) symmetry. On the evidence of our numerical results, the quantum
phase transition between the antiferromagnetic phase and the plaquette phase
is found to be either a direct second order or a very weak first order transition.
\end{abstract}

\maketitle

\selectlanguage{english}

\section{introduction}

In a quantum spin system, the spin order of the classical ground state can
be melted by a quantum fluctuation at zero temperature \cite{Anderson73}, and
the resulting so-called quantum spin liquid has been suggested as a possible
parent state of high temperature superconductivity upon electron or hole
doping \cite{Anderson87}. The quantum fluctuation is usually enhanced with
small spin, low dimensionality, and geometric frustration. Indeed, so far
almost all the candidates of quantum spin liquid discovered or proposed are
effective spin-1/2 systems with or without charge fluctuations on triangular
lattice \cite{CuCN1,CuCN2,SbPd1,SbPd2,Misguich1998}, on Kagome lattice \cite%
{ZnCu1,ZnCu2,BaCu,CuV,Huse2010}, or on honeycomb lattice \cite%
{wangfa2010,sondhi2011,lauchli2011,reuther2011,Meng2010}. Besides the exotic
quantum spin liquid phases, strong quantum fluctuations in spin-1/2 systems
can also lead to highly unconventional quantum critical points. For example,
it has been proposed that a generic direct second order quantum phase
transition between a magnetic ordered phase and a paramagnetic phase with
broken lattice symmetry can exist in spin-1/2 quantum magnets \cite%
{Senthil2004A,Senthil2004B}. Such a transition is forbidden by the classic
Landau-Ginzburg-Wilson-Fisher paradigm, and is called the deconfined quantum
critical point, as it is described by fractionalized quantities instead of
physical order parameters. In the last few years this theoretical proposal
has gained strong numerical evidence by quantum Monte Carlo simulation on
spin-1/2 models with both nearest neighbor Heisenberg coupling and four-spin
interactions \cite{sandvik1,sandvik2,kaulmelko}.

In this paper we will address the question: Can quantum fluctuation lead to
such exotic phases and phase transitions in systems with larger spins in two
dimensions without geometric frustration? Theoretically, this question is
highly nontrivial as the Affleck-Kennedy-Lieb-Tasaki type of valence bond
state (VBS) only exists for spin-1 systems in one dimension. Recent
experiments on the spin-1 magnet Ba$_{3}$NiSb$_{2}$O$_{9}$ suggested that a
highly nontrivial quantum disordered ground state of a two dimensional
spin-1 system is indeed possible \cite{banisbo}.

By using the state of the art tensor renormalization group method \cite%
{Jiang08,Zhao10,Vidal07}, we present strong numerical evidences
for the quantum fluctuation driven exotic physics in the spin-1
bilinear-biquadratic Heisenberg model on the honeycomb lattice. An
interesting but surprising result we find is that all dipole and
quadruple magnetic orders vanish in a phase where the ground state
is staggered quadrupolar ordered in the classical limit, instead
the system develops a translation symmetry breaking plaquette
order. Moreover,
we demonstrate that the transition between the plaquette and
antiferromagnetic AF order is either a direct second order transition or a
very weak first order transition. If it is indeed a direct second order
transition, then it is most likely a deconfined quantum critical point,
which is the first example of deconfined quantum critical point in spin-1
system.

The spin-1 bilinear-biquadratic Heisenberg model reads
\begin{equation}
H=\sum_{\left\langle i,j\right\rangle }\left[ \left( \cos \theta \right)
\mathbf{S}_{i}\cdot \mathbf{S}_{j}+\left( \sin \theta \right) \left( \mathbf{%
\ S}_{i}\cdot \mathbf{S}_{j}\right) ^{2}\right] .  \label{eq:model}
\end{equation}%
The honeycomb lattice has the smallest coordination number in two dimensions
and the effect of quantum fluctuation is the strongest. This model contains
a number of special points. The point $\theta =0$ is the conventional SU(2)
AF Heisenberg model. When $\theta =\pi /4$, $5\pi /4$, or $\pm \pi /2$, the
Hamiltonian is SU(3) invariant, possessing a symmetry higher than the spin
SU(2) symmetry. At $\theta _{\pm }=\pm \arctan 2$, the Hamiltonian can be
expressed purely using the quadrupolar tensor operator

\begin{eqnarray}
\mathbf{Q}_{i} & = & \left(
\begin{array}{c}
S_{ix}^{2}-S_{iy}^{2} \\
\sqrt{3}S_{iz}^{2}-2/\sqrt{3} \\
S_{ix}S_{iy}+S_{iy}S_{ix} \\
S_{iy}S_{iz}+S_{iz}S_{iy} \\
S_{ix}S_{iz}+S_{iz}S_{ix}%
\end{array}%
\right),  \label{eq:QP}
\end{eqnarray}
as
\begin{equation}
H=\sum_{\left\langle i,j\right\rangle }\left( \frac{\sin \theta _{\pm }}{2}%
\mathbf{Q}_{i}\cdot \mathbf{Q}_{j}+\frac{4}{3}\sin \theta _{\pm }\right) .
\end{equation}%
Like the ferromagnetic (FM) spin operator, the uniform quadrupolar operator,
$\mathbf{Q}=\sum_{i}\mathbf{Q}_{i}$, commutes with this Hamiltonian.
However, the staggered quadruple operator, $\mathbf{Q}_{s}=\sum_{i}(-)^{i}%
\mathbf{Q}_{i}$, does not commute with the Hamiltonian.

The two terms in Eq.~(\ref{eq:model}) introduce competition between
different kinds of magnetic orders. The first term favors the conventional
ferromagnetic or antiferromagnetic order, while the second term favors a ferro- or
antiferro-quadrupolar order. This competition causes a strong quantum
fluctuation, especially in the regime $\sin \theta >0$ where the Marshall
sign rule is not applicable to the ground state wave function and the
quantum Monte Carlo suffers the minus-sign problem.

Aspects of the spin-1 bilinear biquadratic model have been explored
previously in the literature. In one dimension, the ground state phase
diagram has been characterized by numerical density matrix renormalization
group method. For $-\pi /4<\theta <\pi /4$, the model gives rise to the
Haldane spin gapped phase, while the ground state for $\pi /4<\theta <\pi /2$
corresponds to a quantum critical phase with power-law spin and quadrupolar
correlations \cite{Fath-Solyom1991,Xiang93}.

\begin{figure}[tbp]
\includegraphics[width=7cm]{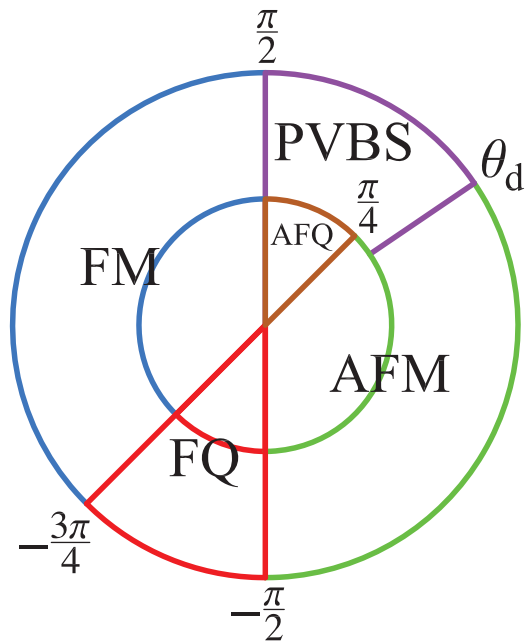}
\caption{(color online) The ground state phase diagram of the spin-1
bilinear-biquadratic Heisenberg model on the honeycomb lattice. The inner
circle is the phase diagram in the classical limit, while the outer circle
is for the corresponding quantum spin model. FM, AFM, FQ, AFQ and PVBS stand for
ferromagnetic, antiferromagnetic, ferro-quadrupolar, antiferro-quadrupolar,
and plaquette valence bond solid phases, respectively. $\protect\theta %
_{d}\approx 0.19\protect\pi $.}
\label{fig:phaseDiagram}
\end{figure}

In two dimensions, the ground state phase diagram has not been firmly
established. In the classical limit, this model possesses four phases \cite%
{Chen-Levy,Papanicolaou88}, as depicted by the inner circle of Fig.~\ref%
{fig:phaseDiagram}. In the lower half plane of $\theta $, the quantum Monte
Carlo simulation\cite{Harada01} and other calculations\cite{lauchli2011}
confirmed the classical phase diagram on square or triangular lattices. In
the upper half plane of $\theta $, there is no quantum Monte Carlo study on
this model due to the minus-sign problem. Other calculations based on mean
field theory and exact diagonalization showed that the phase $\pi /4<\theta
<\pi /2$ is antiferro-quadrupolar ordered on the triangular or square lattice%
\cite{Penc06}.

\section{methods}

\begin{figure}[tbp]
\includegraphics[width=8cm]{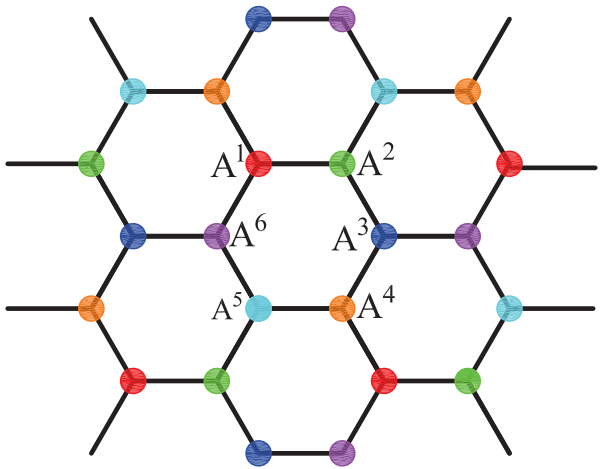}
\caption{(color online) Diagrammatic representation of the tensor-network
wave function on the honeycomb lattice. Tensor $A^{i}$ defined on each
lattice site contains three virtual bond indices and one physical index. }
\label{fig:TNS}
\end{figure}

The tensor renormalization group method recently developed is an accurate
numerical method for studying the ground state of quantum lattice models in
two dimensions\cite{Jiang08,Zhao10,Vidal07}. It does not have the minus-sign
problem encountered in the quantum Monte Carlo simulation and can be used to
study the phase diagram in the whole parameter space. We assume that the
ground state is described by the following tensor-product wave function
\begin{equation}
\left| \Psi \right\rangle =\text{Tr}\prod_{\left\{ i\right\}
}A_{x_{i}y_{i}z_{i}}^{i}\left[ m_{i}\right] \left| m_{i}\right\rangle ,
\end{equation}%
where $m_{i}$ is the eigenvalue of spin operator $S_{i}^{z}$. $%
A_{x_{i}y_{i}z_{i}}^{i}\left[ m_{i}\right] $ is the third-order tensors
defined on the 6 sublattices, as shown in Fig.~\ref{fig:TNS}. The trace is
to sum over all spin configurations and all virtual bond variables. This
wave function satisfies the area law of entanglement entropy. It is an
accurate representation of the ground state wave function. Its accuracy is
determined by the bond dimension $D$. It approaches the exact result in the
limit $D\rightarrow \infty $.

The ground state wave function, or the local tensors $A^{i}$, is determined
by applying the projection operator $\exp \left( -\tau H\right) $ to an
arbitrary initial state $\left| \Psi \right\rangle $ iteratively until it is
converged. Since this model only contains nearest neighbor interactions, $%
\exp \left( -\tau H\right) $ can be divided into a sequence of local
two-site operators approximately by the Trotter-Suzuki decomposition for a
sufficiently small $\tau $. We apply the first order Trotter-Suzuki
decomposition here. In our calculation, we start the projection with a
relatively large $\tau =0.2$ and then reduce it gradually to $10^{-4}$ until
the wave function is converged. In order to find the true ground state and
not being trapped in a local minimum, we start the projection from variety
of possible magnetically ordered states or valence bond solid states. We
choose the converged state which has the lowest energy as the ground state
wave function. A detailed introduction to this method can be found from
Refs.~[\onlinecite{Jiang08,Zhao10}]. This method is a fast and accurate way
to get the ground state wave function.

After obtaining the ground state wave function $\left| \Psi \right\rangle $,
we can evaluate the expectation value of physical variable $O$
\begin{equation}
\left\langle O\right\rangle =\frac{\left\langle \Psi \right| O\left| \Psi
\right\rangle }{\left\langle \Psi |\Psi \right\rangle }.
\label{eq:expectValue}
\end{equation}%
By contracting the physical indices, both $\left\langle \Psi \right| O\left|
\Psi \right\rangle $ and $\left\langle \Psi |\Psi \right\rangle $ can be
also expressed as tensor network. The contraction of tensors is
achieved by computing the dominant eigenvector of the corresponding one
dimensional transfer matrix using the infinite time-evolving block
decimation (iTEBD) method\cite{Vidal07} beyond unitary evolution. The iTEBD is also an iterative
projection method and the truncation error does not accumulate during the
iteration. The largest eigenvector of the transfer matrix is represented by
a matrix product state with bond dimension $\chi $, which determines the
accuracy of the expectation values.

In our calculations, we found that the ground state energy is converged when
the bond dimension $D\geq 12$, while the expectation values of physical
variables become stable when the parameter $\chi \geq 30$ (see Fig.~\ref%
{fig:E_D}). Thus, we choose $D=12$ and $\chi =30$ throughout the
calculations.

\begin{figure}[tbp]
\includegraphics[width=8cm]{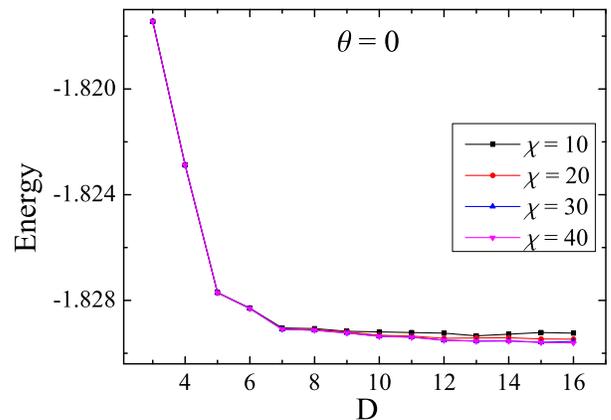}
\caption{(color online) The ground state energy density as a function of $D$
with different values of $\protect\chi$ at $\protect\theta = 0$. The curves
of $\protect\chi=30$ and $\protect\chi=40$ are almost on top of each other. }
\label{fig:E_D}
\end{figure}

\section{results}

\begin{figure}[tbp]
\includegraphics[width=8cm]{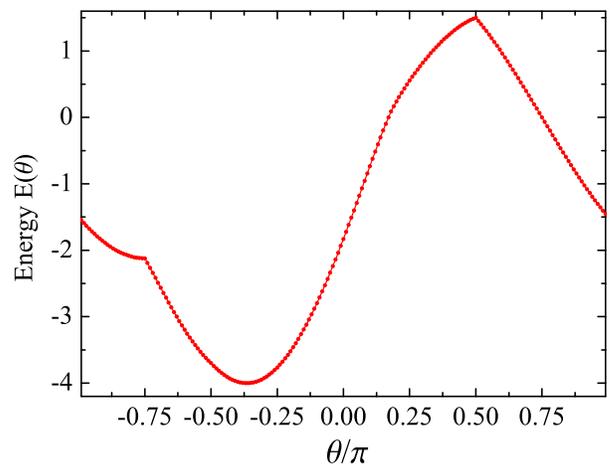}
\caption{(color online) The ground state energy density as a function of $%
\protect\theta $.}
\label{fig:E_theta}
\end{figure}

The ground state energy shows that most part of the quantum phase diagram
matches with the classical phase diagram. Fig.~\ref{fig:E_theta} displays
the $\theta $-dependence of the ground state energy and Fig.~\ref%
{fig:dE_theta} displays its first and second derivative. The first
derivative is calculated with Hellmann-Feynman theorem, which is more
accurate than numerical differentiation from the ground state energy. We
find that there are four phase transitions, located at $\theta =-3\pi /4$, $%
\pm \pi /2$ and $\theta _{d}\approx 0.19\pi $, respectively. Among them, $%
\theta =-3\pi /4$ and $\pm\pi /2$ are first order transitions. The
transition at $\theta _{d} $ is a second order one. This transition point is
shifted below the classical value $\theta =\pi /4$, which will be discussed
later on.

\begin{figure}[tbp]
\includegraphics[width=8cm]{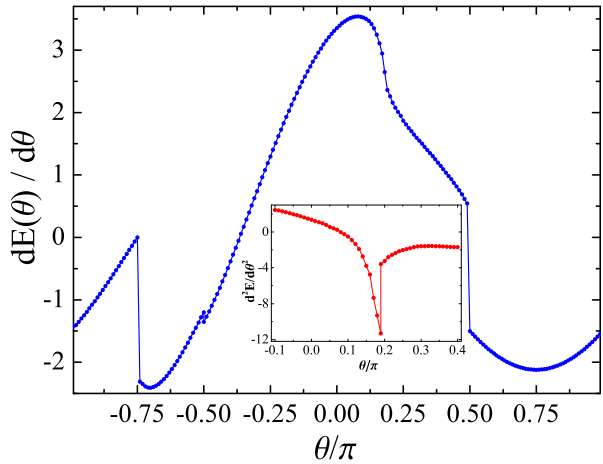}
\caption{(color online) The first and second (the inset)
derivative of the ground state energy with respect to $\protect\theta$.}
\label{fig:dE_theta}
\end{figure}

\begin{figure}[tbp]
\includegraphics[width=8cm]{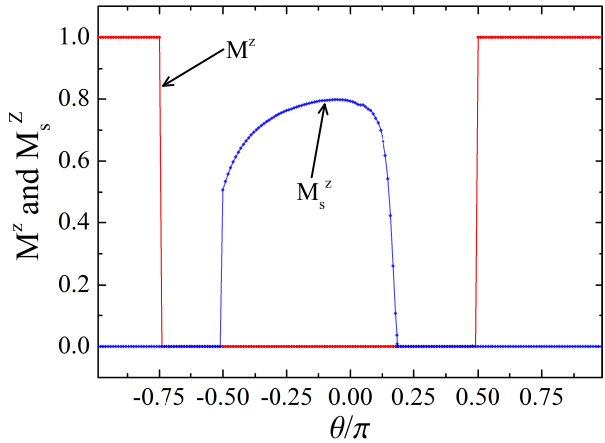}
\caption{(color online) Uniform (red line) and staggered (blue line)
magnetization per site as a function of $\protect\theta $.}
\label{fig:MzMsz_theta}
\end{figure}

\begin{figure}[tbp]
\includegraphics[width=8cm]{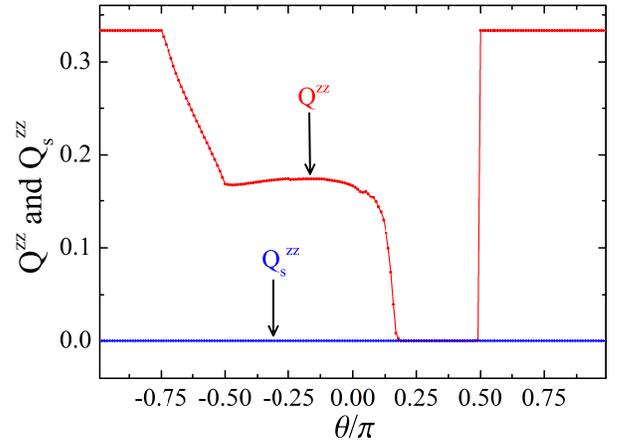}
\caption{(color online) Uniform (red line) and staggered (blue line) spin
quadruple moment per site as a function of $\protect\theta $. }
\label{fig:QzzQszz_theta}
\end{figure}

To clarify the phase diagram, we calculate various order parameters, i.e.
the magnetization

\begin{equation}
M^{z}=\sum_{i}\langle S_{iz}\rangle ,
\end{equation}
the staggered magnetization
\begin{equation}
M_{s}^{z}=\sum_{i}(-)^{i}\langle S_{iz}\rangle ,
\end{equation}
the ferro-quadrupolar moment
\begin{equation}
Q^{zz}=\sum_{i}\langle S_{iz}^{2}\rangle -2/3 ,
\end{equation}
and the antiferro-quadrupolar moment
\begin{equation}
Q_{s}^{zz}=\sum_{i}(-)^{i}\langle S_{iz}^{2}\rangle ,
\end{equation}
in the four phases, respectively. Fig.~\ref{fig:MzMsz_theta} shows the $%
\theta $ -dependence of $M^{z}$ and $M_{s}^{z}$. The ground state is found
to have FM long range order for $\pi /2<\theta <5\pi /4$, and AF long range
order for $-\pi /2<\theta <\theta _{d} $. In these phases, the quadruple
moment $Q^{zz}$ is finite. In the region $-3\pi /4<\theta <-\pi /2$, both
magnetization and staggered magnetization vanish, however, the quadrupolar
moment is finite, shown in Fig. \ref{fig:QzzQszz_theta}. It corresponds to a
ferro-quadrupolar phase, in agreement with both the semiclassical\cite%
{Chen-Levy,Papanicolaou88} and quantum Monte Carlo\cite{Harada01} results.

In the staggered magnetization curve a sharp jump shows at $\theta _{c}=-\pi
/2$. This feature was also observed in the quantum Monte Carlo calculation
on a square lattice\cite{Harada01}. But the quadrupolar moment is finite and
changes continuously at this point. So a first order phase transition
occurs, in consistence with the conclusion drawn from the first derivative
of the ground state energy.

As expected, the staggered quadrupolar moment $Q_{s}^{zz}$ vanishes in the
FM, AF and ferro-quadrupolar phases. A surprising result is that this moment
also vanishes in the classical staggered quadrupolar phase $\pi /4<\theta
<\pi /2$, i.e. the quantum fluctuation suppresses completely the staggered
quadrupolar order, different from the previous studies on the triangular or
square lattices \cite{Penc06}. More interestingly, the critical point has
been shifted by the quantum fluctuation from $\pi /4$ to about $0.19\pi $,
which excludes the SU(3) AF Heisenberg spin-1 model from any long-range
magnetic order.

\begin{figure}[tbp]
\includegraphics[width=8cm]{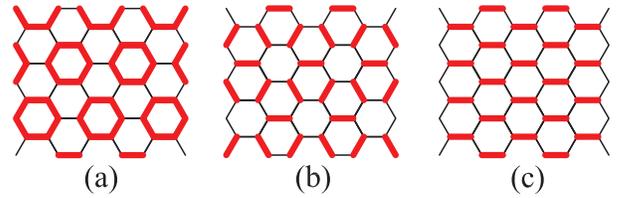}
\caption{(color online) Pictorial representation of three possible VBS
patterns considered in calculation. (a) plaquette; (b) columnar; (c)
staggered. The red thicker bonds represent stronger correlation while the
black thinner bonds represent weaker correlation. }
\label{fig:VBSpattern}
\end{figure}

To further characterize the phase for $\theta _{d} <\theta <\pi /2$, we have
performed a thorough exploration of three possible VBS patterns (Fig.~\ref%
{fig:VBSpattern}) on the honeycomb lattice. It has been checked that
whatever VBS patterns we start with, it always converges into the plaquette
order phase (Fig.~\ref{fig:VBSpattern}(a)) under renormalization group flow.
Hence the ground state energy of the plaquette VBS phase is the lowest. This
plaquette order phase explicitly breaks the lattice translation symmetry,
but not the spin SU(2) symmetry. A naive picture of this plaquette order is
that, in order to minimize the ground state energy, the spins on one third
of the minimal hexagons of the honeycomb lattice form the VBS phase, like
the Haldane gapped phase in one dimension.

\begin{figure}[tbp]
\includegraphics[width=8cm]{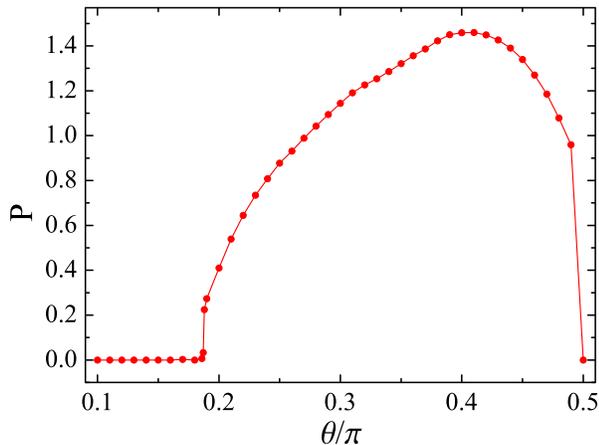}
\caption{(color online) The plaquette order parameter as a function of $%
\protect\theta $. }
\label{fig:PVBS_theta}
\end{figure}

In order to detect the plaquette order, we calculate the plaquette order parameter defined as
\begin{equation}
P=\frac{\sum_{\left\langle i,j\right\rangle \in red}\langle \mathbf{S}%
_{i}\cdot \mathbf{S}_{j}\rangle }{2\sum_{\left\langle i,j\right\rangle \in
black}\langle \mathbf{S}_{i}\cdot \mathbf{S}_{j}\rangle }-1,
\end{equation}%
where $\left\langle i,j\right\rangle \in red$ $\left( black\right) $ means
the two nearest neighbor spins connected by red (black) bond of Fig.~\ref%
{fig:VBSpattern}(a). Fig.~\ref{fig:PVBS_theta} shows the $\theta $%
-dependence of the plaquette order parameters. Both the plaquette and AF
orders vanish simultaneously and continuously at the critical point
$\theta _{d}$. 
This observation suggests that this plaquette-AF transition is in
fact a second order transition, in consistence with the conclusion drawn
from the first and second derivative of the ground state energy with respect
to $\theta $ in Fig.~\ref{fig:dE_theta}. But we are still unable to rule out
the possibility of a very weak first order transition, partly due to the
finite bond dimension $D$.

However, around the transition point between antiferromagnetic and plaquette valence bond solid phases,
we have evaluated the ground state wave function using the cluster update approach
proposed by Wang et al.\cite{cluster_update11}. The cluster update considers
long range entanglement by computing larger block size. The accuracy of the
computation near a second order phase transition can be improved by using
relatively small cluster size (6 sites).

\section{discussion and conclusion}

A second order transition between AF and VBS order was originally predicted
theoretically as the deconfined quantum critical point. This theory is based
on the observation that the topological defect (Skyrmion) of the AF order
parameter carries a finite lattice momentum, thus after the AF order is
suppressed by the Skyrmion proliferation, the system automatically enters
the VBS order. The previous studies on deconfined criticality were focused
on spin-1/2 systems only, and this theory has gained strong numerical
evidence from quantum Monte Carlo simulation on a spin-1/2 model on the
square lattice with both two-body and four-body interactions \cite%
{sandvik1,sandvik2,kaulmelko}. Our present result actually gives rise to a
 possible deconfined quantum critical point in the spin-1 systems.
Using the techniques in Ref.~[\onlinecite{haldane,sachdev1990}], we can show
that for spin-1 systems on the honeycomb lattice the momenta carried by the
Skyrmion will precisely lead to the plaquette order pattern after the
Skyrmion proliferation \cite{note}.

Due to the critical point between AF and plaquette order at $\theta
_{d} $, our numerical result also implies that the ground
state of the SU(3) AF Heisenberg model on the honeycomb lattice has a
plaquette order. This result concurs with the recent studies on the SU(N)
Heisenberg model \cite{hermele2010}, which suggested that the SU(N) spins
tend to form block singlets that are commensurate with the lattice.

To summarize, the ground state phase diagram of the quantum spin-1
bilinear-biquadratic Heisenberg model on a honeycomb lattice has been
determined precisely. Besides the ferromagnetic, antiferromagnetic and ferro-quadrupolar phases,
a plaquette order phase is found in the region of $\theta_{d} <\theta <\pi /2$
for the first time, where the classical AF or staggered quadrupolar order is
completely suppressed by quantum fluctuations. The quantum phase transition
between AF and the plaquette order phase is found to be either a direct second order or a very weak
first order transition. This is a possible candidate of a
deconfined quantum critical point in a quantum spin-1 system.
Further investigation on the critical properties around this point is desired.

We would like to thank Ling Wang for stimulating discussion. This work is
supported by NSFC and the grants of National Program for Basic Research of
MOST of China. Cenke Xu is supported by the Sloan Research Fellowship.


\begin{thebibliography}{99}
\bibitem{note} Another columnar VBS order (Fig.~\ref{fig:VBSpattern}(b)) has
the same symmetry as the plaquette order, which in principle can also be a
candidate phase of the Skyrmion condensate. However, in this part of the
phase diagram the energy of the columnar VBS order is higher than the
plaquette order. One can see this by noting that the SU(3) invariant point
is in this plaquette phase, and the plaquette phase can be adiabatically
connected to a SU(3) singlet phase, while the VBS order in Fig.~\ref%
{fig:VBSpattern}(b) cannot. Thus we conclude that the plaquette order is
energetically more favorable than the columnar VBS order close to the SU(3)
point.
\end{thebibliography}
\end{document}